\pdfoutput=1
\documentclass{optica-article}

\journal{opticajournal} 
\articletype{Research Article}

\usepackage{amsmath}
\usepackage{graphicx}
\usepackage{booktabs}
\usepackage{siunitx}
\usepackage{bm}
\usepackage{xcolor}
\usepackage{soul}
\usepackage[normalem]{ulem}
\soulregister\cite7
\soulregister\textit7

\begin{document}

\title{Photonic reservoir computing with complex networks}

\author{Sion Park,\authormark{1} Kohei Watabe,\authormark{1} Satoshi Sunada,\authormark{2} Tomoki Yamagami,\authormark{1} and Atsushi Uchida\authormark{1,*}}
\address{\authormark{1}Department of Information and Computer Sciences, Saitama University, 255 Shimo-okubo, Sakura-ku, Saitama City, Saitama 338-8570, Japan}
\address{\authormark{2}Faculty of Mechanical Engineering, Institute of Science and Engineering, Kanazawa University, Kakuma-machi, Kanazawa, Ishikawa, 920-1192, Japan}
\email{\authormark{*}auchida@mail.saitama-u.ac.jp}

\begin{abstract*}
Photonic reservoir computing has attracted increasing attention as a fast and low-cost approach for time-series prediction. Photonic reservoir computing utilizes the high speed, broad bandwidth, and spatial parallelism of light. However, the effect of the internal connection structure (network topology) on the computing performance has not been investigated for large-scale photonic reservoirs. In this study, we experimentally and numerically demonstrate photonic reservoir computing using a spatial light modulator to systematically evaluate the relationship between the network topology and the performance of reservoir computing. We introduce complex network structures such as small-world and scale-free network topologies of the internal nodes in the reservoir. We perform the memory capacity measurement and the one-step-ahead prediction task of the chaotic time series to compare the performance. We found that the small-world network exhibits the maximum memory capacity and the best prediction performance. Our numerical calculations reveal that the performance of the time-series prediction can be optimized by changing the rewiring probability of the network and the leak rate of the reservoir. We also implement photonic human brain network as a reservoir, which is designed by the connectomes of human brain activities. We found that the network topology strongly affects the performance of reservoir computing, and the small-world network structure outperforms the other configurations.
\end{abstract*}


\section{Introduction}
The amount of time-series data with temporal variations, such as audio, video, biological signals, and financial data, has increased owing to the spread of Internet of Things (IoT) devices and the advancement of sensing technologies. Techniques for accurately predicting such data are essential in various research fields, including autonomous driving, anomaly detection, and natural language processing. Conventional feedforward neural networks have difficulty handling the temporal context and sequential relationships among inputs. To overcome this limitation, recurrent neural networks (RNNs) have been proposed and widely used \cite{Rumelhart1986}. RNNs feature recurrent connections between nodes in the hidden layer. Their loop structure retains past input information as internal states and integrates it with the current input, enabling information processing according to the context of the time series \cite{Elman1990}. However, RNNs are generally trained using backpropagation through time, which often suffers from vanishing or exploding gradient problem as the time series becomes longer \cite{Bengio1994}. Consequently, it is difficult to learn long-term dependencies, and the iterative update of network weights requires significant computational cost and time.

Reservoir computing has been proposed as a machine learning technique to alleviate these challenges \cite{Jaeger2001,Jaeger2002,Maass2002,Jaeger2004}. Reservoir computing consists of three layers: an input layer, a reservoir layer, and an output layer. The connection weights from the input layer to the reservoir and the internal connection weights within the reservoir are randomly fixed, whereas only the connection weights of the output layer are trained using linear regression. Therefore, reservoir computing has the advantage of faster training and significantly reduced computational cost compared to conventional RNNs. Furthermore, photonic reservoir computing, which utilizes the characteristics of light for the reservoir, has been intensively investigated in recent years because it leverages the high speed, broad bandwidth, and spatial parallelism of light \cite{Larger2012,Brunner2013,Nakayama2016,Bueno2017,Kuriki2018,Takano2018,Sunada2019,Sugano2020,Sunada2020,Porte_2021,Sunada2021,Kanno2022,Hasegawa2023}. Among photonic reservoir computing approaches, spatially multiplexed implementations using spatial light modulators (SLMs) have been extensively investigated as a means of exploiting the intrinsic spatial parallelism of light \cite{Bueno:18,Antonik2019,Rafayelyan2020,Antonik2020JSTQE}. Previous studies have demonstrated large-scale spatial photonic reservoirs and recurrent photonic neural networks with thousands to tens of thousands of optical nodes for applications including reinforcement learning, human-action recognition, image classification, and spatiotemporal information processing \cite{Bueno:18,Antonik2019,Rafayelyan2020,Antonik2020JSTQE}. These studies have established SLM-based free-space optical architectures as an important platform for large-scale photonic reservoir computing.

From the viewpoint of complex-network science, various reservoir network topologies, including small-world, scale-free, random, and other structured networks, have been investigated \cite{Kawai2017,Kitayama2022,Kawai2019,Haluszczynski2019,Dale2021}. Kawai \textit{et al.} systematically investigated small-world echo-state networks and showed that a small-world topology can enhance the echo-state property and signal propagation, particularly when the input and output nodes are spatially segregated \cite{Kawai2019}. Haluszczynski and R\"ath compared Erd\H{o}s--R\'enyi, small-world, and scale-free reservoir networks and showed that their relative prediction performance depends on the target dynamical system \cite{Haluszczynski2019}. Other studies have also demonstrated that reservoir connectivity and topology influence reservoir dynamics, while more complex structures are not necessarily superior under all conditions \cite{Dale2021}. These previous studies indicate that network topology can influence reservoir performance, while the advantageous topology depends on the reservoir configuration, target system, and computational task. In addition, most systematic comparisons of complex-network topology have been performed using numerical reservoir models, whereas corresponding experimental studies using large-scale photonic reservoirs remain limited.

In this study, we experimentally implement and systematically compare these network topologies on a 1024-node SLM-based photonic reservoir platform. By keeping the physical platform and reservoir size unchanged, we directly evaluate the influence of the internal network topology on memory capacity and chaotic time-series prediction. Furthermore, we perform corresponding numerical simulations and directly compare the topology-dependent performance with the experimental results. The main contribution of this study is therefore the systematic physical implementation and comparison of different complex-network topologies in a common large-scale photonic reservoir, together with a direct experiment--simulation comparison of their topology-dependent computational performance. We further investigate the joint dependence of the computational performance on the network topology and reservoir dynamical parameters and evaluate a human-brain functional network as a reservoir topology.

In the present SLM-based implementation, each macropixel (a group of pixels) on the SLM is regarded as a reservoir node, and the main role of the optical system is to perform nonlinear transformation and state generation for a large number of spatially multiplexed reservoir nodes in parallel. This optical spatial parallelism enables 1024 physical reservoir nodes to be simultaneously represented and optically transformed on the same SLM platform, although the network coupling and readout are digitally implemented in the present system.

Experimental control of reservoir topology has also recently been demonstrated in a photonic time-delay reservoir using multiple optical feedback loops, where the preferred virtual-node topology was shown to depend on the computational task \cite{Zhang2026Topology}. In contrast, the present work focuses on the systematic implementation and comparison of complex-network structures, including regular, small-world, random, and scale-free networks, in a large-scale spatially multiplexed photonic reservoir.

This paper is organized as follows. Section 2 describes the scheme and training method of reservoir computing. Section 3 explains the experimental setup for photonic reservoir computing using an SLM and the generation method of the network topologies. Section 4 shows the experimental and numerical results, and Section 5 discusses the relationship between the topology and the prediction performance based on numerical calculations. Section 6 describes the photonic human brain network. Section 7 shows discussions. Finally, Section 8 concludes this paper and discusses future works.

\begin{figure}[htbp]
\centering
\includegraphics[width=0.6\linewidth]{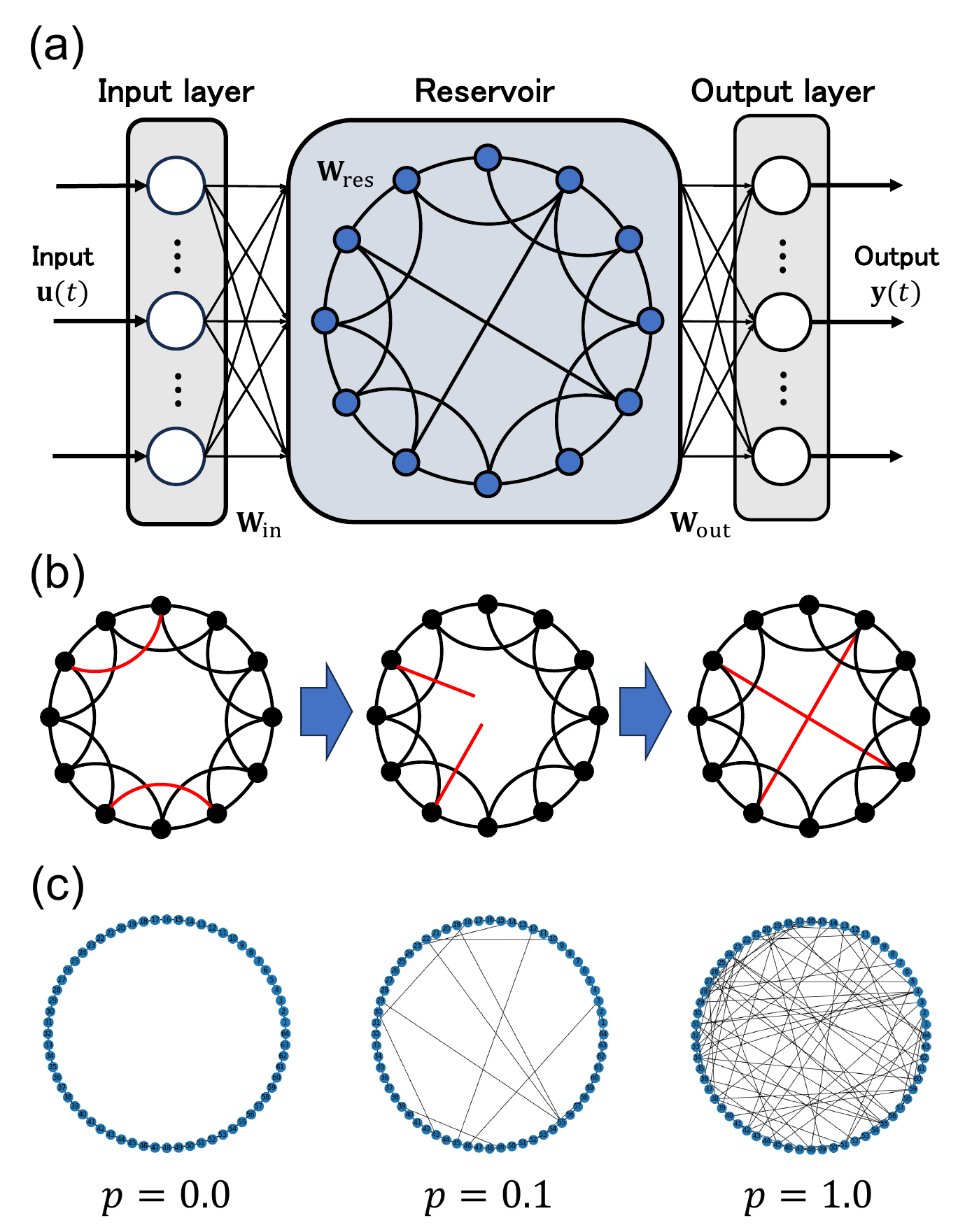}
\caption{Reservoir computing using complex networks. (a) Schematic diagram. (b) Generation method of the Watts-Strogatz model. (c) Examples of networks generated by the Watts-Strogatz model depending on the rewiring probability $p$.}
\label{fig:fig1}
\end{figure}

\section{Photonic reservoir computing}
Figure~\ref{fig:fig1}(a) shows the schematic diagram of reservoir computing, consisting of an input layer, a reservoir layer, and an output layer. The input signal is distributed to each reservoir node with input weights and nonlinearly transformed within the reservoir. In this study, we introduce complex networks into the connection structure of the reservoir. 

Figure~\ref{fig:fig1}(b) shows the generation method of the WS model. First, a regular network connected to neighboring nodes is generated based on the average degree $k$. Next, an edge to be rewired is selected based on the rewiring probability $p$, and the reconnected edge is randomly changed, where self-loops and multiple edges are avoided. The network structure is determined by the value of $p$. As examples of Fig. 1(c), a regular network is found at $p = 0.0$, a small-world network is generated at around $p = 0.1$, and a random network is obtained at $p = 1.0$.

In the training phase, the node state vectors of the reservoir corresponding to the inputs are acquired, and the weights of the output layer are calculated by using ridge regression. In the testing phase, the output signal is obtained by a linear combination of the node states and the trained output weights.

In our SLM-based photonic reservoir computing system, each macropixel (a group of pixels) on the SLM is regarded as a reservoir node. The signal (pixel value) transmitted to each macropixel is calculated on a computer. We use the adjacency matrix obtained from the complex network model as the internal connection structure of the reservoir, which is treated as the connection weights (i.e., the connection matrix). The signal calculated on the computer is sent to the SLM, and a nonlinear transformation is induced by the modulation of the SLM. The optical intensity after the transformation is measured by a camera, and the optical intensity of each macropixel is utilized as the node state.

\subsection{Input layer}
We explain the model of each layer for photonic reservoir computing in the following. Let $\mathbf{u}(t) = (u_1(t), \dots, u_{N_u}(t))^T$ be the input signal at time $t$. The input signal is distributed to each reservoir node by a randomly generated input weight matrix $\mathbf{W}_{in} \in \mathbb{R}^{N_x \times N_u}$, where $N_u$ and $N_x$ are the number of input dimensions and the number of reservoir nodes, respectively. The elements of $\mathbf{W}_{in}$ are given by uniform random numbers in the interval \numrange{-1}{1}.

\subsection{Reservoir}
The input signals are mixed through recurrent connections in the reservoir. In the SLM-based photonic reservoir computing, the node states are calculated on a computer and sent to the SLM, where the nonlinear transformation of the node states is applied on the SLM. The node states are updated by the model equation in the experiment and are expressed as follows.

\begin{equation}
I(t+1) = F(\mathbf{W}_{res}\mathbf{x}(t) + \mathbf{W}_{in}\mathbf{u}(t))
\label{eq:exp_nonlinear}
\end{equation}
\begin{equation}
\mathbf{x}(t+1) = (1 - \alpha)\mathbf{x}(t) + \alpha I'(t+1)
\label{eq:exp_update}
\end{equation}
The update of the node states is performed through the following procedure. The linear combination of the current node state $\mathbf{x}(t)$ and the input $\mathbf{u}(t)$ is calculated using the reservoir connection weight matrix $\mathbf{W}_{res}$ and the input weight matrix $\mathbf{W}_{in}$, respectively. The calculated result is sent to the SLM, and the optical intensity $I(t+1)$ subjected to the nonlinear transformation by the intensity modulation of the SLM is obtained, where $F(x) = \cos x$. The obtained optical intensity $I(t+1)$ is normalized to the range of $[\pi, 2\pi]$, which is denoted as $I'(t+1)$, and $\mathbf{x}(t)$ is updated using the leak rate $\alpha$ and $I'(t+1)$. Here, the leak rate $\alpha$ is a parameter that controls the memory of the past states. A smaller $\alpha$ results in a slower change in $\mathbf{x}(t)$, whereas a larger $\alpha$ strongly reflects the new input and the nonlinear response.

The update equation of the node states in the numerical calculation is expressed as follows. 

\begin{equation}
\mathbf{x}(t+1) = (1 - \alpha)\mathbf{x}(t) + \alpha \cos(\mathbf{W}_{res}\mathbf{x}(t) + \mathbf{W}_{in}\mathbf{u}(t))
\label{eq:num_update}
\end{equation}

We approximate the nonlinear response of the SLM with a cosine function and reproduce the same procedure as the experiment (i.e., linear combination, nonlinear transformation, and state update with the leak rate) on a computer. This enables us to validate the experimental results and understand the tendencies through parameter changes.

\subsection{Output layer}
The output signal $\mathbf{y}(t) \in \mathbb{R}^{N_y}$ is defined by the weighted linear combination of the reservoir state $\mathbf{x}(t)$ and the output weight matrix $\mathbf{W}_{out} \in \mathbb{R}^{N_y \times N_x}$ as follows.

\begin{equation}
\mathbf{y}(t) = \mathbf{W}_{out} \mathbf{x}(t)
\label{eq:output}
\end{equation}
where $N_y$ is the number of output dimensions. In the training phase, $\mathbf{W}_{out}$ is estimated to minimize the output error with respect to the target signal $\mathbf{d}(t)$. In this study, we calculate $\mathbf{W}_{out}$ using ridge regression to suppress overfitting and ensure numerical stability.

\begin{figure}[htbp]
\centering
\includegraphics[width=\linewidth]{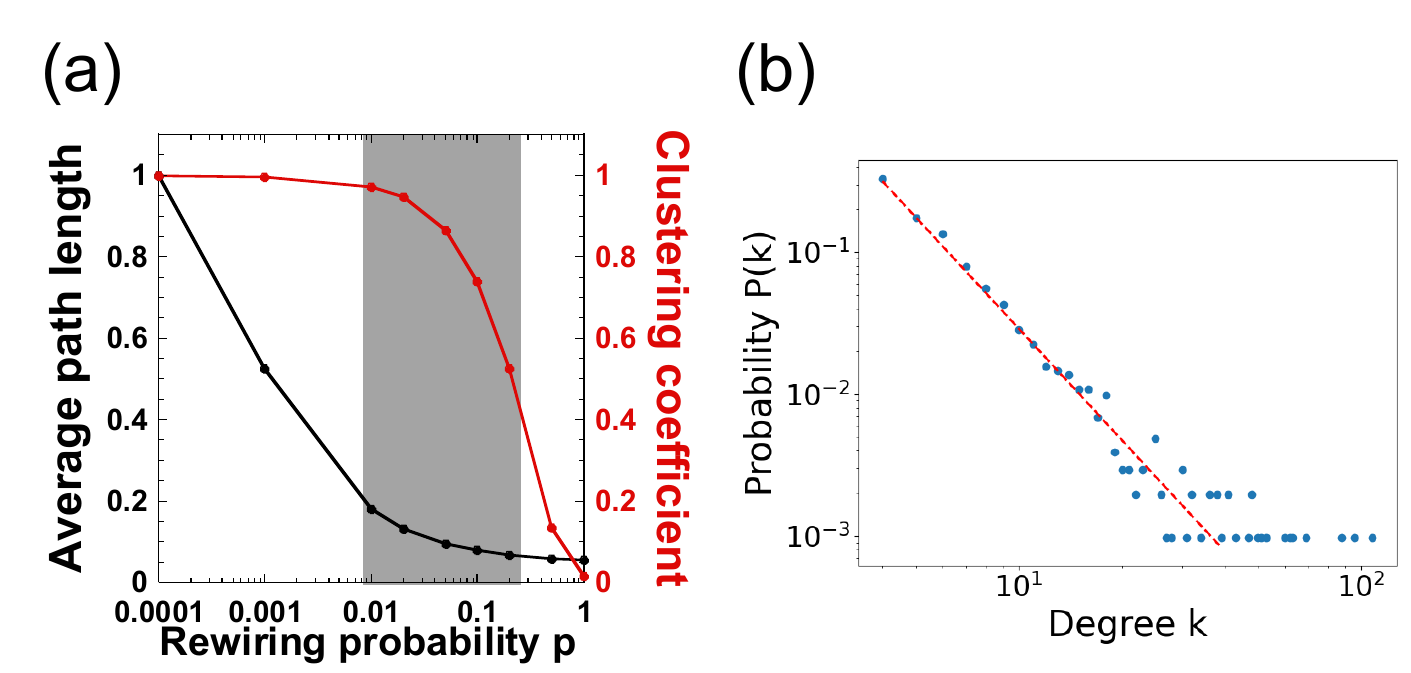}
\caption{Evaluation of generated complex networks. (a) Clustering coefficient and the average path length of the Watts-Strogatz (WS) model with 1024 nodes. (b) Degree distribution of the network generated by the Barab\'{a}si-Albert (BA) model with 1024 nodes.}
\label{fig:fig2}
\end{figure}

\subsection{Generation and evaluation of complex networks}
In this study, we use the adjacency matrix of a complex network as the internal connection structure ($\mathbf{W}_{res}$) of the reservoir. We introduce two types of complex network models.

\def\labelenumi{(\theenumi)}
\begin{enumerate}
    \item Watts-Strogatz (WS) model (small-world network)~\cite{Watts1998} \par
    In the WS model, a regular network (regular ring lattice) is constructed, and each edge is rewired with a probability $p$ to generate a network with small-world properties. In this study, we use a WS network with 1024 nodes. Figure \ref{fig:fig2}(a) shows the changes in the average path length and the clustering coefficient when the rewiring probability $p$ is changed for the WS model with 1024 nodes. We define the range indicated in gray ($0.01 \le p \le 0.2$) as the small-world network region, where the characteristics of a small-world network show a large clustering coefficient and a short average path length.
    
    \item Barab\'{a}si–Albert (BA) model (scale-free network)~\cite{Barabasi1999} \par
    The Barab\'{a}si–Albert (BA) model generates a scale-free network. Nodes are added one by one, and more new nodes are attached to existing nodes with higher degrees. As a result, a large number of low-degree nodes and a small number of high-degree nodes (called hubs) emerge, and a power-law degree distribution is observed in the BA network. We use the number of edges of a new node, $m$, as a generation parameter. Figure \ref{fig:fig2}(b) shows the degree distribution of the BA model with 1024 nodes. The degree distribution of the network follows a power law with a power exponent of 2.61.
\end{enumerate}

\begin{table}[htbp]
\centering
\caption{Comparison of network features between the Watts-Strogatz (WS) and Barabási-Albert (BA) models with $N_x=1024$.}
\label{tab:topology_stats}
\begin{tabular}{lcccc} 
\toprule
Topology & Average degree & Clustering coefficient & Average path length & Diameter \\
\midrule
WS model & 8.00 & 0.47 & 5.00 & 9.0 \\
BA model & 7.97 & 0.04 & 3.16 & 5.0 \\
\bottomrule
\end{tabular}
\end{table}



We evaluate the complex networks generated by the two models using network features. The network diameter is defined as the maximum shortest-path length between any pair of nodes in the network and is used as a global measure characterizing the maximum topological distance between nodes and the overall extent of connectivity across the network. Table ~\ref{tab:topology_stats} shows the evaluation results of the network features with 1024 nodes for each model. Here, the parameters of the WS model are the rewiring probability $p = 0.1$ and the average degree $k = 8$. The parameter of the BA model is the number of edges of a new node, $m = 4$. Comparing between the two models, the WS model exhibits a larger clustering coefficient, whereas the BA model exhibits a shorter average path length.

\begin{figure}[htbp]
\centering
\includegraphics[width=\linewidth]{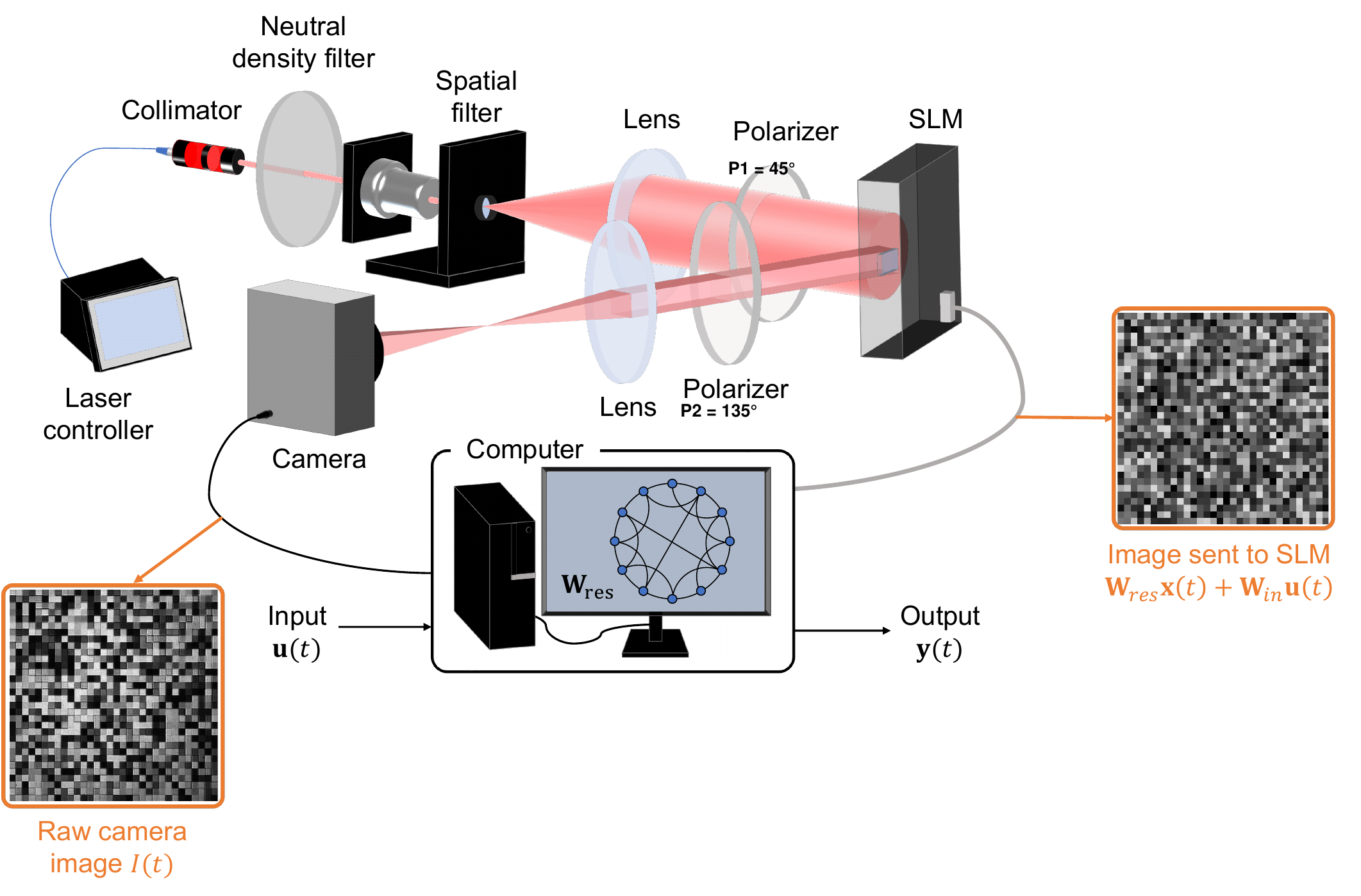}
\caption{Experimental setup of the SLM-based photonic reservoir computing. SLM: spatial light modulator. The polarization angles are defined with respect to the liquid-crystal alignment direction of the SLM (0$^\circ$): P1 = 45$^\circ$ and P2 = 135$^\circ$, forming a crossed-polarizer configuration for phase-to-intensity conversion.}
\label{fig:fig3}
\end{figure}

\section{Experimental methods}
\subsection{Experimental setup}
Figure~\ref{fig:fig3} shows the experimental setup used for photonic reservoir computing. The experimental setup consists of a semiconductor laser, a collimator, a neutral density (ND) filter, a spatial filter, a lens, a polarization filter, an SLM, a CMOS camera, and a control computer. The output beam of a semiconductor laser (LP642-PF20, Thorlabs) is collimated, and the light intensity is adjusted by the ND filter. The laser beam is spatially expanded through the spatial filter and injected into the SLM (X13138-01, $1272 \times 1024$ pixels, \qty{60}{frames/s}; Hamamatsu Photonics). The light is phase-modulated by the SLM, and is converted into intensity modulation by passing through the polarization filter. Its intensity pattern is measured by the CMOS camera with \num{12}-bit quantization (C11440-36U, $1920 \times 1200$ pixels, \qty{64.9}{frames/s}; Hamamatsu Photonics). The optical intensity is acquired from the captured image, and the SLM input pattern (feedback signal) at the next time step is calculated by a personal computer (PC) (CPU: Intel Core i9-14900, \qty{2.00}{GHz}; RAM: \qty{16.0}{GB}; OS: Windows 11). The calculated signal is sent to the SLM with \num{8}-bit quantization to generate optical spatiotemporal dynamics~\cite{Pecora2014,Morijiri2023}. Four least significant bits of the camera image are discarded.

The pixels on the SLM are grouped into multiple square regions to form macropixels, and each macropixel is treated as a reservoir node~\cite{Morijiri2023}. In the experiment, we use $512 \times 512$ pixels at the center of the SLM. By setting one macropixel to $16 \times 16$ pixels, we construct a reservoir with $32 \times 32 = 1024$ nodes in total. When the node state from the camera image is obtained, the optical intensity corresponding to each macropixel region is extracted, and the average optical intensity of the pixels included in the macropixel is used as the state value of the node. In the following experiments, we fix the number of nodes at 1024 and change the network topology to compare the performance while the effect of the scale difference is excluded.

We use a polarization system to convert the phase modulation into intensity modulation for the readout. The liquid-crystal alignment direction of the SLM is defined as 0$^\circ$. The polarization direction of the incident light is set to 45$^\circ$ with respect to this direction by the first polarizer (P1), while the second polarizer (P2), serving as an analyzer, is set to 135$^\circ$. Thus, P1 and P2 form a crossed-polarizer configuration. The relative phase modulation produced by the SLM is thereby converted into intensity modulation, and the intensity observed by the camera exhibits a sinusoidal nonlinear response depending on the pixel values applied to the SLM.

Based on the experimentally obtained intensity modulation characteristics of the SLM, the pixel value $I_{CAM}(t)$ of the macropixel acquired by the camera at time $t$ is modeled as follows.

\begin{equation}
I_{CAM}(t) = A \cos(2\pi f I_{SLM}(t) - \theta) + B
\label{eq:icam_model}
\end{equation}
Here, $I_{SLM}(t)$ is the pixel value input to the SLM at time $t$. $A$, $\theta$, and $B$ represent the amplitude of the intensity modulation, the initial phase, and the offset, respectively. $f$ is the frequency of the phase modulation with respect to the input pixel value. The average intensity $I_{SLM}(t)$ of the macropixel on the SLM is converted into that of the CMOS camera $I_{CAM}(t)$ by magnifying the 2D beam pattern.

Next, we describe the feedback process to the system. The image acquired by the camera is sent to the computer, and the pixel values within each macropixel are averaged. Then, the feedback strength $\beta$ is multiplied and a modulo operation with a correction coefficient $a$ is applied. The mapping of the feedback system is expressed as follows.

\begin{equation}
I_{CAM}(t+1) = A \cos \{ 2\pi f \cdot (\beta(I_{CAM}(t) - \theta) \bmod a) \} + B
\label{eq:feedback_map}
\end{equation}
Equation~\eqref{eq:feedback_map} indicates a mapping of a cosine function. This nonlinearity results in complex optical spatiotemporal dynamics.
From the experimentally obtained intensity modulation characteristics, we obtain $A = 116$, $f = 1/207$, and $B = 120$, and we confirm that a sufficient modulation range can be obtained for the input. In addition, we introduce $a$ as a coefficient to correct the input range corresponding to one period of the intensity modulation (a phase change of $2\pi$), because a phase shift is observed and the intensity does not become maximum at an input of 0. 

The initial phase $\theta$ and the coefficient $a$ are not described in the numerical simulations, and the model equation of the intensity modulation is expressed as follows.

\begin{equation}
I_{CAM}(t+1) = A \cos(2\pi f \cdot \beta I_{CAM}(t)) + B
\label{eq:icam_num}
\end{equation}

To perform reservoir computing with network connections, the input signal $u(t)$ and the network connections from other nodes are added to $I_{CAM}(t)$ with the weight matrices, as described in Eq. (\ref{eq:num_update}).  

We explicitly distinguish the optical and digital operations in the present implementation. The optical part consists of the spatially multiplexed representation of the 1024 reservoir nodes on the SLM, the nonlinear optical intensity transformation produced by the SLM and polarization optics, and the parallel optical generation of the spatial node states detected by the camera. In contrast, the complex-network topology and the corresponding reservoir connection matrix are generated digitally. The linear combination of the current reservoir states and input signals, the calculation of the feedback signal for the next time step, data acquisition and normalization, and ridge-regression training and readout are also performed digitally on the computer. Therefore, the present system is a hybrid optical--digital reservoir rather than a fully optical reservoir-computing system.

In addition, the nonlinear response of the optical system is not identical among the SLM macropixels (reservoir nodes), and node-dependent variations occur in the amplitude and phase of the nonlinear response. To compensate for these variations, an additional calibration procedure is applied in the experiment so that the nonlinear responses are approximately uniform across the reservoir nodes. In contrast, the numerical simulation assumes an identical nonlinear response function for all reservoir nodes and therefore does not require this calibration.

\subsection{Generation of spatiotemporal dynamics}
In our SLM-based photonic reservoir computing, optical spatiotemporal dynamics are generated by repeatedly updating the SLM input pattern at the next time step using the optical intensity measured by the camera. The intensity for each macropixel is extracted from the camera image at time $t$, and the next feedback signal is calculated using the normalized state values. By repeating this process, the dynamics is expanded in both the spatial and temporal domains. These optical spatiotemporal dynamics are used for generating high-dimensional node states.

\begin{figure}[htbp]
\centering
\includegraphics[width=0.9\linewidth]{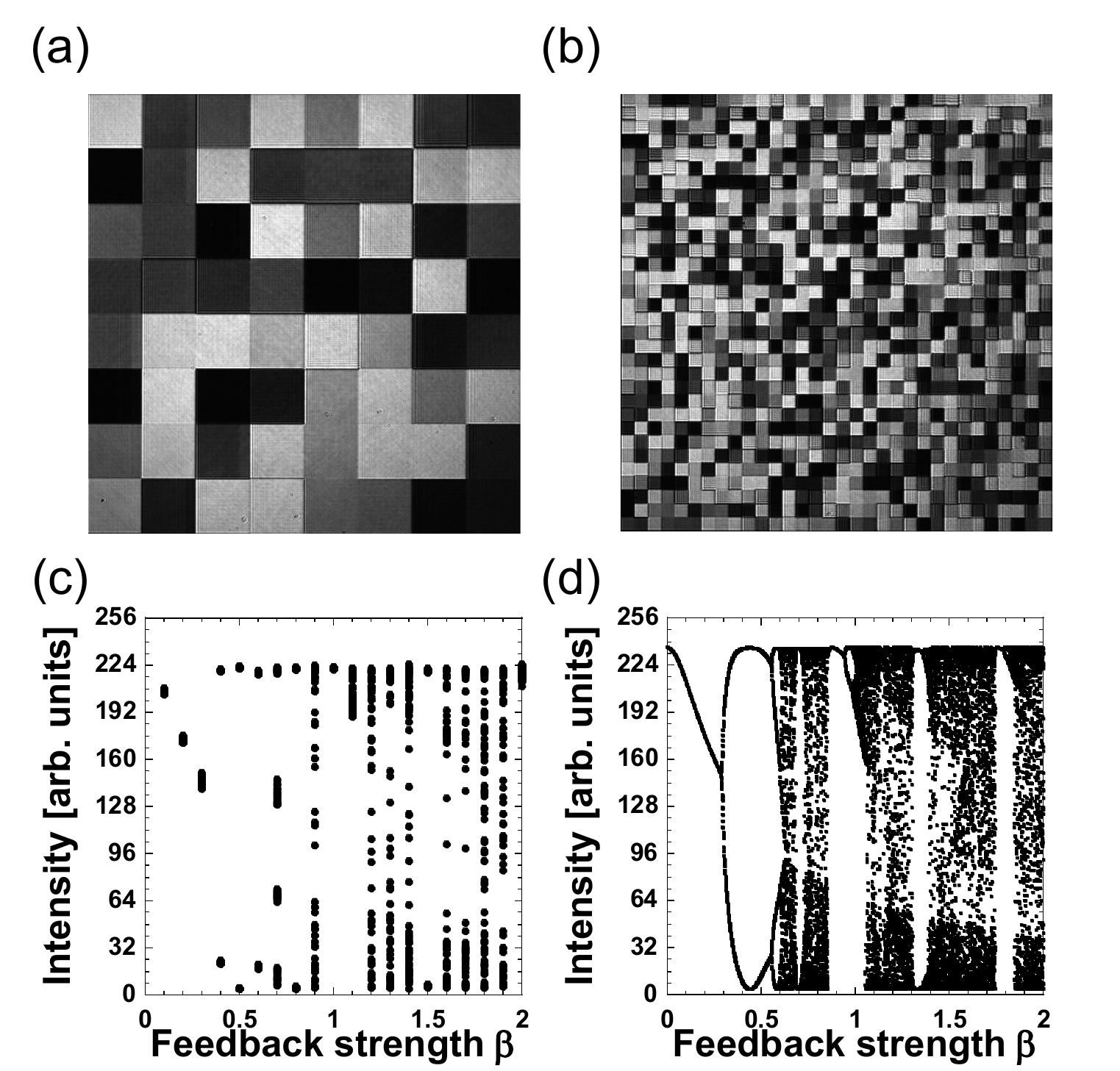}
\caption{Spatiotemporal dynamics generated by SLM with (a) 64 and (b) 1024 macropixels (nodes) without network connections. Bifurcation diagrams as a function of the feedback strength $\beta$ obtained from (c) the experiment and (d) the numerical simulation.}
\label{fig:fig4}
\end{figure}

We generate spatiotemporal dynamics with 64 ($8 \times 8$) and 1024 ($32 \times 32$) macropixels to confirm the effect of the spatial resolution (number of nodes) on the representation of spatiotemporal states. Figures ~\ref{fig:fig4}(a) and \ref{fig:fig4}(b) show the light intensity of 64 and 1024 nodes without network connections, respectively. We confirm that the diversity of states is poor owing to the limited spatial degrees of freedom for 64 nodes. On the contrary, for 1024 nodes, rich intensity variations appear in the spatial domain, and a high-dimensional state representation accompanied by complex spatiotemporal dynamics is obtained. In the following experiments, we use the 1024-node configuration, which is expected to improve computing performance due to the large-scale state representation.

An important feature of this implementation is that the nonlinear transformation of these spatial node states is performed optically and in parallel across the SLM. Thus, increasing the number of macropixels increases the number of spatially multiplexed physical reservoir nodes that undergo the optical nonlinear transformation in parallel. This spatial parallelism is a principal hardware advantage of the SLM-based photonic implementation.

Next, we change the feedback strength $\beta$ to systematically evaluate the effect of $\beta$ on the dynamics. Figures ~\ref{fig:fig4}(c) and \ref{fig:fig4}(d) show the bifurcation diagrams obtained from the experiment and the numerical simulation under the same conditions without network connections, respectively. In both cases, as $\beta$ increases, a route from a steady state to a periodic oscillation appears, and a chaotic oscillation through a period-doubling route to chaos is observed. Therefore, we confirm that the experimental result agrees with the numerical result.

The computational performance of reservoir computing generally depends on the dynamical operating regime, and a favorable operating regime can vary depending on the computational task and reservoir parameters~\cite{Nakayama2016,Bueno2017}. As shown in Fig. 4(d), the first transition from periodic dynamics through period doubling to chaotic dynamics occurs around $\beta = 0.6$ in the numerical simulation. In the present study, we set $\beta = 1.0$ as an operating point within the nonlinear dynamical regime. This operating point provides sufficiently rich reservoir dynamics for the subsequent computational experiments. The bifurcation diagrams are used to characterize how the reservoir dynamics changes with $\beta$. We perform the performance evaluation under this common operating condition in both the experiment and numerical simulations.

We define the internal connections of the reservoir using the adjacency matrix generated from complex network models. We use a binary adjacency matrix of 0s and 1s for the reservoir connection matrix $\mathbf{W}_{res}$, where 1 and 0 indicate the presence and absence of the connection, respectively (see Eq. (3)). This enables us to eliminate the effects of weights and directionality, and clearly evaluate the effect of the network topology on the computing performance. This binary-weight condition is employed to isolate the effect of network topology from additional effects arising from the sign and magnitude distributions of the connection weights. We note that the WS model at $p=1$ provides a random-network baseline under the same binary-weight condition, enabling a comparison between the small-world and random-network reservoirs without introducing differences in the weight distribution. For comparison, we use (i) the WS model with different rewiring probabilities $p$ and (ii) the BA model. The WS model can continuously change its structure from a regular network ($p=0$), a small-world network, and a random network ($p=1$).

\subsection{Tasks for performance evaluation}
We use (1) a memory capacity (MC) measurement~\cite{Jaeger2002} and (2) a one-step-ahead prediction task of the Mackey-Glass chaotic time series~\cite{Mackey1977,Bueno2017} to evaluate the performance of reservoir computing from different newtork topologies. In both tasks, we train the output weights using ridge regression. We set the regularization coefficient (ridge parameter) to $\lambda = 2.0 \times 10^{-9}$ \cite{Kitayama2022}.

\def\labelenumi{(\theenumi)}
\begin{enumerate}
    \item In the MC measurement~\cite{Jaeger2002}, we use uniform random numbers in the interval \numrange{-1}{1} as the input signal $u(n)$, and we train the reservoir using the past input $u(n - k)$ delayed by $k$ steps as the target signal. The correlation between the target and prediction signals is calculated for each delay, and their sum is defined as the MC. The calculation formulas for the MC are expressed as follows.
    \begin{equation}
    m(k) = \frac{\mathrm{cov}^2[u(n - k), y_k(n)]}{\mathrm{var}[u(n)] \mathrm{var}[y_k(n)]}
    \label{eq:mc_degree}
    \end{equation}
    \begin{equation}
    MC = \sum_{k=1}^{\infty} m(k)
    \label{eq:mc_total}
    \end{equation}
    Here, $y_k(n)$ is the output signal for $d(n) = u(n - k)$, $\mathrm{cov}[x, y]$ is the covariance of $x$ and $y$, and $\mathrm{var}[x]$ is the variance of $x$. A higher MC indicates a higher memory effect. We set the conditions as follows: the number of nodes $N_x = 1024$, the number of input dimensions $N_u = 1$, the number of output dimensions $N_y = 20$, where $N_y$ denotes the maximum delay considered in the memory-capacity evaluation. Thus, delays from 1 to 20 discrete time steps are included, and the total memory capacity is calculated by summing the corresponding delay-dependent memory functions. The number of training points is 5000, the number of testing points of 5000, and the leak rate $\alpha = 1.0$.
    
    \item We perform a one-step-ahead prediction of chaotic time series generated by the Mackey-Glass equation with a delay time of $\tau = 17$~\cite{Mackey1977,Bueno2017}. The Mackey--Glass equation is numerically integrated with a time step of 0.10. The generated time series is then downsampled by taking one data point every 50 integration steps, resulting in a sampling interval of 5.0 for the discrete time series used in the one-step-ahead prediction task. The sampling interval is selected such that the change between successive sampled points is sufficiently large to make the one-step-ahead prediction task nontrivial.

We use the normalized mean square error (NMSE) as a performance metric as follows.
    \begin{equation}
    \mathrm{NMSE} = \frac{1}{L} \sum_{n=1}^{L} \frac{(y(n) - y_{tar}(n))^2}{\mathrm{var}(y_{tar})}
    \label{eq:nmse}
    \end{equation}
    Here, $y(n)$ is the output signal, $y_{tar}(n)$ is the target signal, and $\mathrm{var}(y_{tar})$ is the variance of $y_{tar}$. An NMSE closer to 0 indicates a higher prediction performance. We set the conditions as follows: the number of nodes $N_x = 1024$, the number of input dimensions $N_u = 1$, the number of output dimensions $N_y = 1$, the number of training points of 2000, the number of testing points of 2000, and the leak rate $\alpha = 0.7$.
\end{enumerate}

\section{Experimental and numerical results of memory capacity and time-series prediction task}
\begin{figure}[htbp]
\centering
\includegraphics[width=\linewidth]{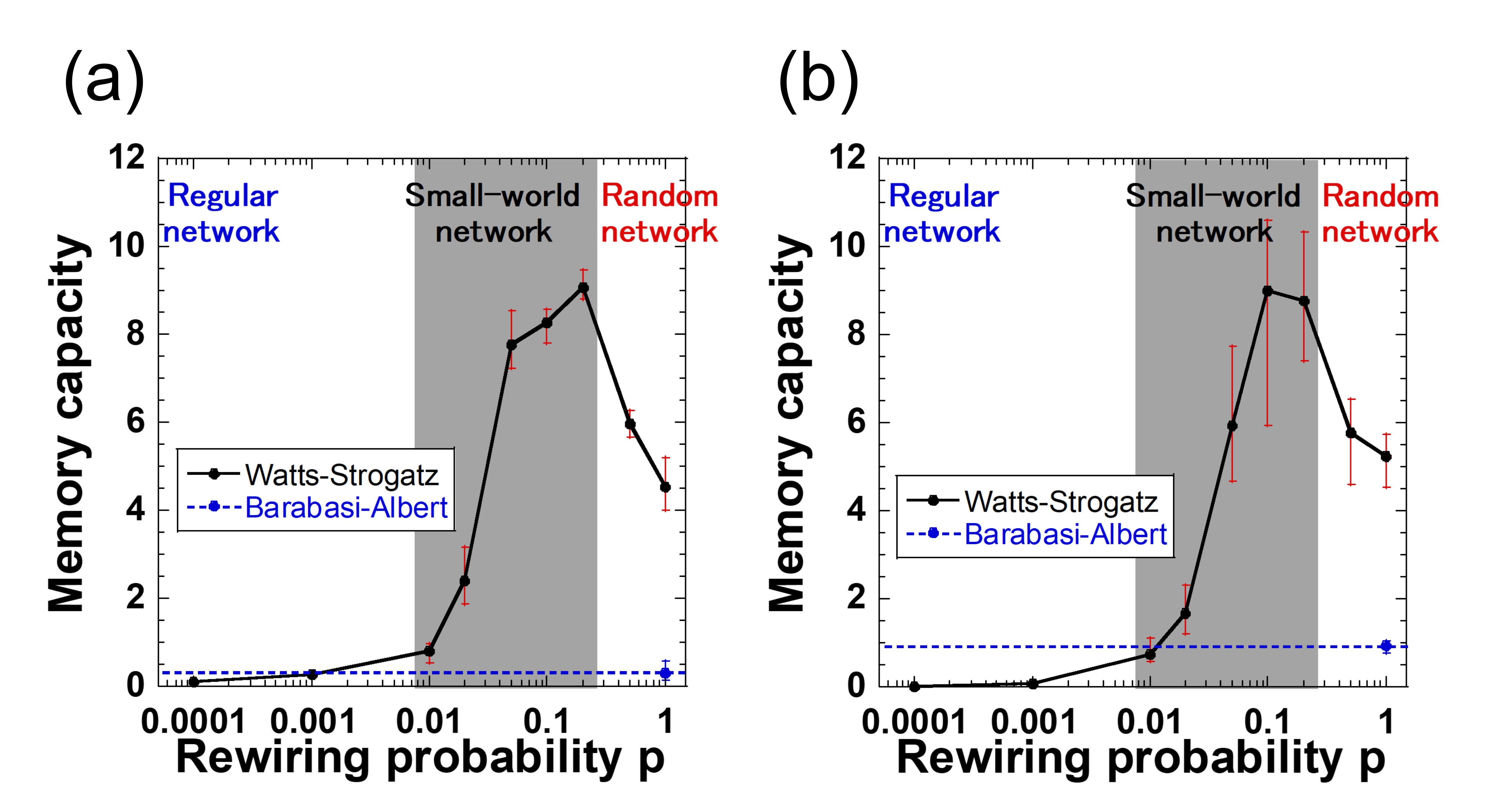}
\caption{(a) Experimental and (b) numerical results of the memory capacity when the rewiring probability $p$ is changed. Solid black line: Watts-Strogatz (WS) model, dotted blue line: Barabási-Albert (BA) model. The error bars indicate the results of three trials in the experiment and ten trials in the numerical simulation.}
\label{fig:fig5}
\end{figure}

In this section, we show both experimental and numerical results using the photonic reservoir computing with complex network. First, we evaluate the memory capacity when the rewiring probability $p$ is varied in the WS model. Figure ~\ref{fig:fig5}(a) shows the experimental results of the memory capacity. At $p = 0.0001$ in a regular network, the memory capacity is very small at $0.10$, indicating that the information of past inputs is not stored in the reservoir. On the contrary, as $p$ increases, the memory capacity increases significantly, and a remarkable improvement is observed in the small-world region ($0.1 \le p \le 0.25$, the gray region in Fig. ~\ref{fig:fig5}(a)). In particular, the memory capacity reaches the maximum value of $MC = 9.08$ at $p = 0.2$. We speculate that this is because the input information propagates efficiently throughout the reservoir while maintaining an appropriate memory capability, which is achieved by introducing an adequate number of shortcut connections and a large clustering effect on the local connections.

In the BA model, we adjust the conditions to have the same connection density as the small-world network. The memory capacity of the BA mode is small ($MC = 0.31$, the dotted blue line in Fig. 5(a)), and the memory capacity is significantly inferior to that of the small-world network. The BA model has a biased degree distribution including hubs, and the information tends to concentrate on specific nodes. Therefore, it is difficult to obtain the distributed node states, which may be required for time-series prediction tasks. From these results, we confirm that the memory capacity of the reservoir strongly depends on the topology in the SLM-based photonic reservoir computing, and memory capacity is maximized in the small-world region of the WS model.

Figure ~\ref{fig:fig5}(b) shows the numerical results of the memory capacity. Compared with the experimental results in Fig. ~\ref{fig:fig5}(a), the overall characteristics are very similar, i.e., the maximum memory capacity is obtained in the small-world region (the gray region in Fig. ~\ref{fig:fig5}(b)). However, the rewiring probability at which the maximum memory capacity is obtained is different ($p = 0.1$). We speculate that this discrepancy may be caused by the influence of noise in the experimental environment.

\begin{figure}[htbp]
\centering
\includegraphics[width=0.9\linewidth]{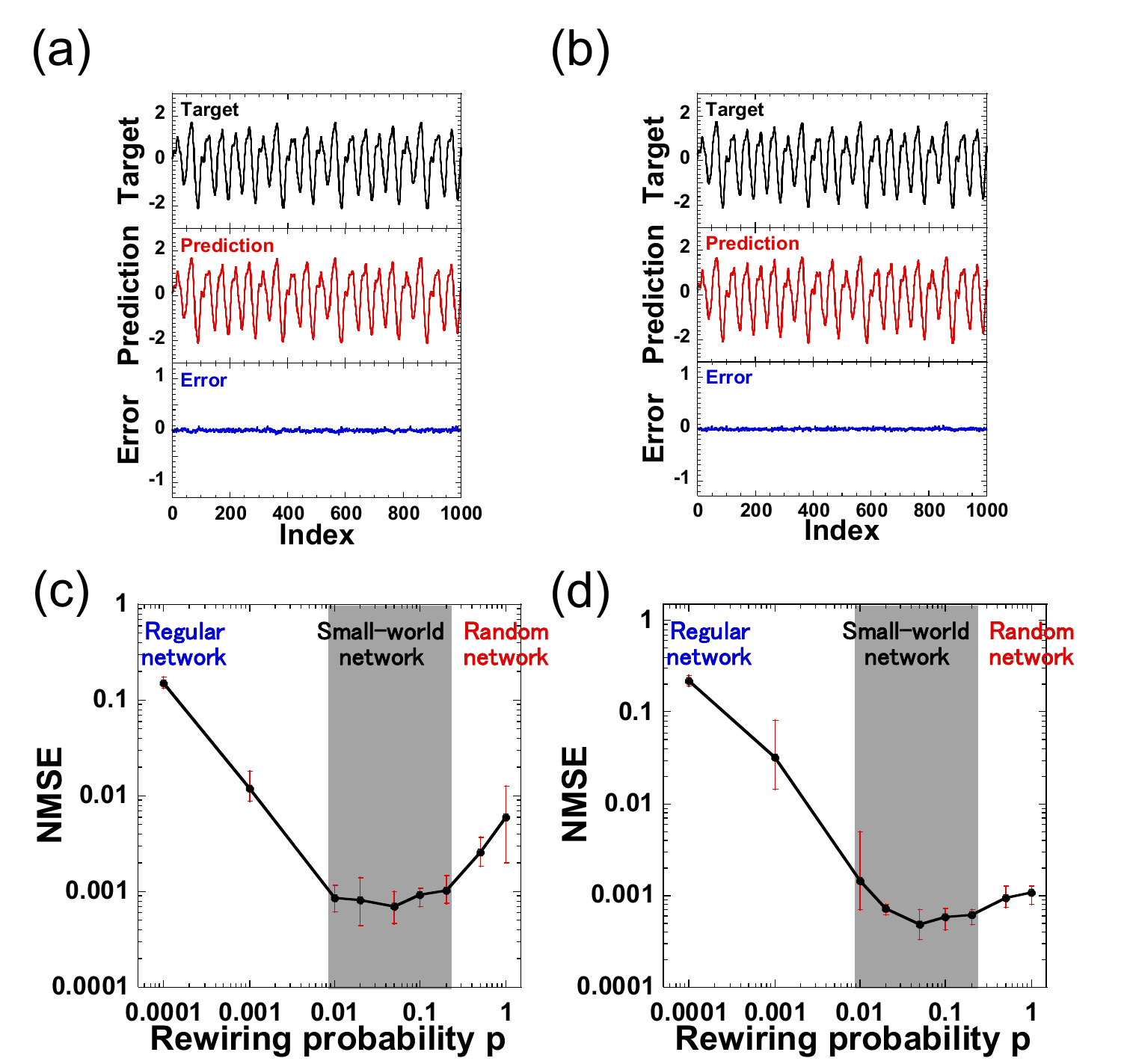}
\caption{Temporal waveforms of (black) the target signal, (red) the prediction signal , and (blue) the error between them when the small-world network ($p = 0.05$) is used for (a) the experiment and (b) the numerical simulation. The NMSEs of the time-series prediction task when the rewiring probability is changed in (c) the experiment and (d) the numerical simulation. (Watts-Strogatz (WS) model: black line). The error bars indicate the results of three trials in the experiment and ten trials in the numerical simulation.}
\label{fig:fig6}
\end{figure}

Next, we evaluate the performance of the one-step-ahead prediction task of the Mackey-Glass chaotic time series. We use NMSE as the prediction performance metric in Eq. (\ref{eq:nmse}), where a smaller value indicates a higher prediction accuracy. Figure~\ref{fig:fig6} shows the experimental and numerical results of the chaotic time-series prediction task. Figure~\ref{fig:fig6}(a) shows an example of time series for $p=0.05$. The target (black) and predicted (red) time series are matched very well, and a small error (blue) is obtained. Figure~\ref{fig:fig6}(b) shows the numerical results, and it is very similar to that of the experimental results in Fig.~\ref{fig:fig6}(a).

Figure~\ref{fig:fig6}(c) shows the experimental result of NMSE as a function of $p$. In the WS model at $p = 0.0001$ (i.e., regular network), a large error is obtained, and the prediction performance is degraded. On the contrary, as $p$ increases, the NMSE decreases rapidly, and the best performance is obtained in the small-world region ($0.01 \le p \le 0.25$). In particular, the NMSE reaches $6.41 \times 10^{-4}$ at $p = 0.05$, and the prediction error is improved in more than two orders of magnitude compared with the regular network. The value of $p$ is further increased to become a random network, and the performance is degraded again. Figure~\ref{fig:fig6}(d) shows the numerical results of the prediction task. Very similar characteristics are obtained to the experimental results, and we verify the validity of the experimental results.

\begin{table}[htbp]
\centering
\caption{Comparison of the performance of the memory capacity (MC) and the time-series prediction task (NMSE) between the WS and BA models.}
\label{tab:results}
\begin{tabular}{lcc} 
\toprule
Model & Memory capacity (MC) & Time-series prediction performance (NMSE) \\
\midrule
WS & 9.08 & $6.41 \times 10^{-4}$ \\
BA & 0.31 & $7.02 \times 10^{-2}$ \\
\bottomrule
\end{tabular}
\end{table}

Table ~\ref{tab:results} summarizes the performance comparison of the memory capacity and the time-series prediction task between the WS and BA models. The BA model exhibits $NMSE = 7.02 \times 10^{-2}$, and a large error is obtained, compared with the best condition of the WS small-world network ($NMSE = 6.41 \times 10^{-4}$). Therefore, we found that the scale-free structure is not effecitve for the prediction task using the SLM-based photonic reservoir computing, and a small-world structure attributes the improvement of the prediction performance.

Both MC and NMSE tend to be improved in the small-world region of the WS model. However, the optimized parameter values do not match between the MC and the time-series prediction task, i.e., MC is maximized at $p = 0.2$ for the MC measurement, and NMSE is minimized at $p = 0.05$ for the time-series prediction task in experiment. Thus, the optimal topology depends on the evaluation task. We consider that this discrepancy is because the MC mainly evaluates the linear delayed reproduction ability, whereas the time-series prediction requires a balance between the nonlinear transformation and the short-term memory capacity. Therefore, for the topology design, it is important to evaluate the performance using different tasks, rather than optimizing it by using only a single task.

\begin{figure}[htbp]
\centering
\includegraphics[width=\linewidth]{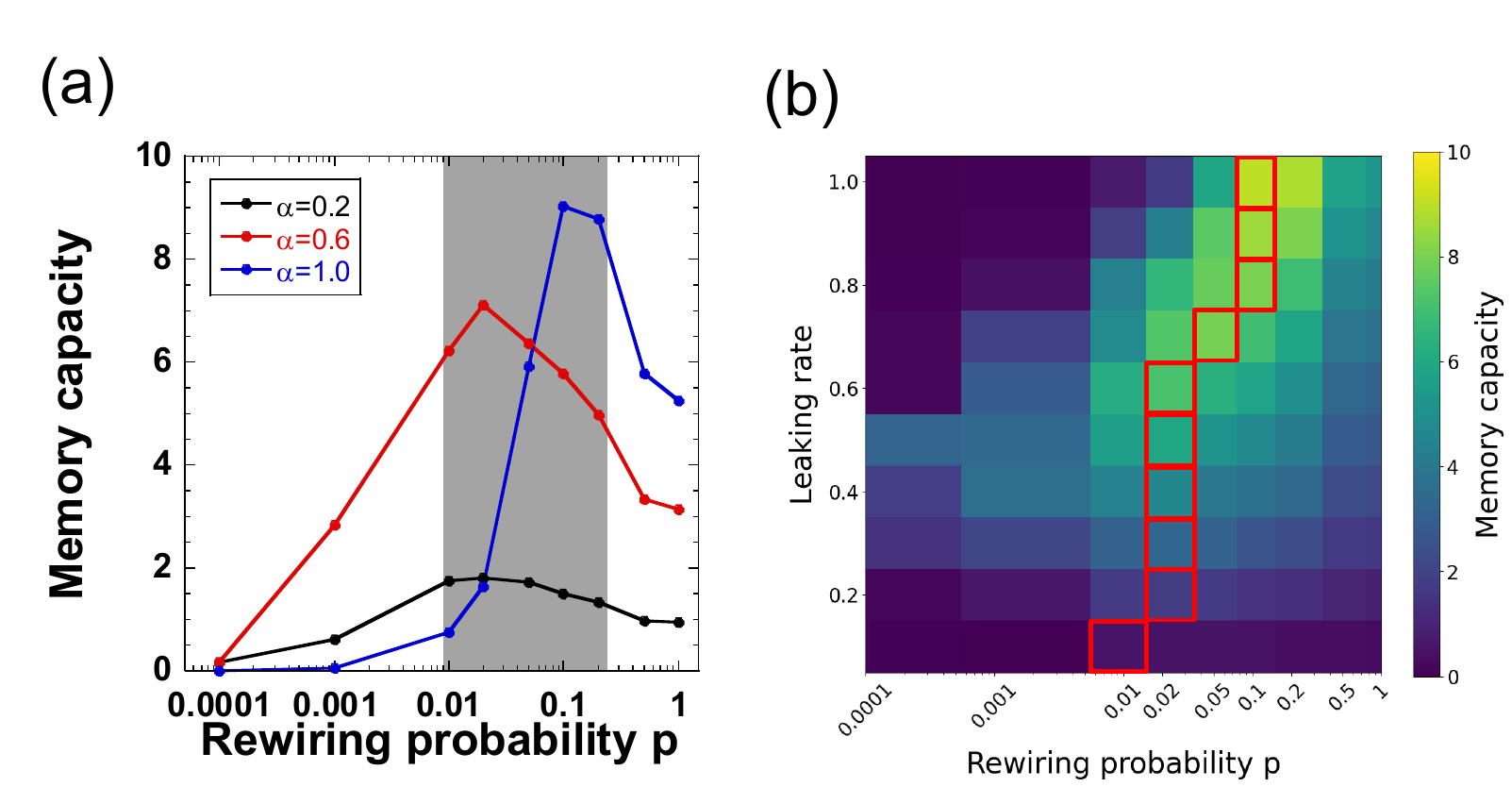}
\caption{Memory capacity (MC) by changing the leak rate $\alpha$ and the rewiring probability $p$. (a) MC when varying $p$ at $\alpha = 0.2$ (black), $\alpha = 0.6$ (red), and $\alpha = 1.0$ (blue). (b) Two-dimensional map of the MC as functions of $\alpha$ and $p$. Red squares indicate the optimal conditions for fixed $\alpha$ by changing $p$.}
\label{fig:fig7}
\end{figure}

\section{Relationship between reservoir network topology and performance}
\subsection{Parameter dependence of the leak rate $\alpha$ and the network topology for memory capacity}
In Section 4, we showed the experimental results agree well with the numerical results. In this section, we systematically investigate the relationship between the network topology of the reservoir and its performance by numerical calculations. We focus on the leak rate $\alpha$, which determines the time scale of the reservoir, and we evaluate the changes in MC and NMSE of the chaotic time-series prediction task when the rewiring probability $p$ is changed in the WS model.

We investigate the effect of the rewiring probability $p$ of the WS model on MC when the leak rate $\alpha$ is varied. The MC is a metric that reflects the retention ability of past inputs, and it is a fundamental measure for evaluating the temporal memory characteristics of the reservoir. Figure ~\ref{fig:fig7}(a) shows the MC by changing $p$ with different $\alpha$. We found a tendency that the MC increases as $\alpha$ increases for a large $p$. We consider this is because the reservoir states are updated more frequently as the increase in the leak rate, and the reservoir states are affected by the input signal more effectively. Figure ~\ref{fig:fig7}(b) shows the two-dimensional map of the MC when $p$ and $\alpha$ are changed simultaneously. For all $\alpha$, the memory capacity reaches its maximum in the small-world region ($0.01 \le p \le 0.2$). Therefore, we confirm that the small-world topology consistently provides a high memory retention ability, compared with the regular network (small $p$) and the random network (large $p$).
However, the optimal parameter value of $\alpha$ for which the memory capacity is maximized depends on $p$. a larger $\alpha$ is required for a larger $p$ (the red squares in Fig. ~\ref{fig:fig7}(b)). This is because more network connections are required to maintain the memory capacity for a larger $\alpha$.

\begin{figure}[htbp]
\centering
\includegraphics[width=\linewidth]{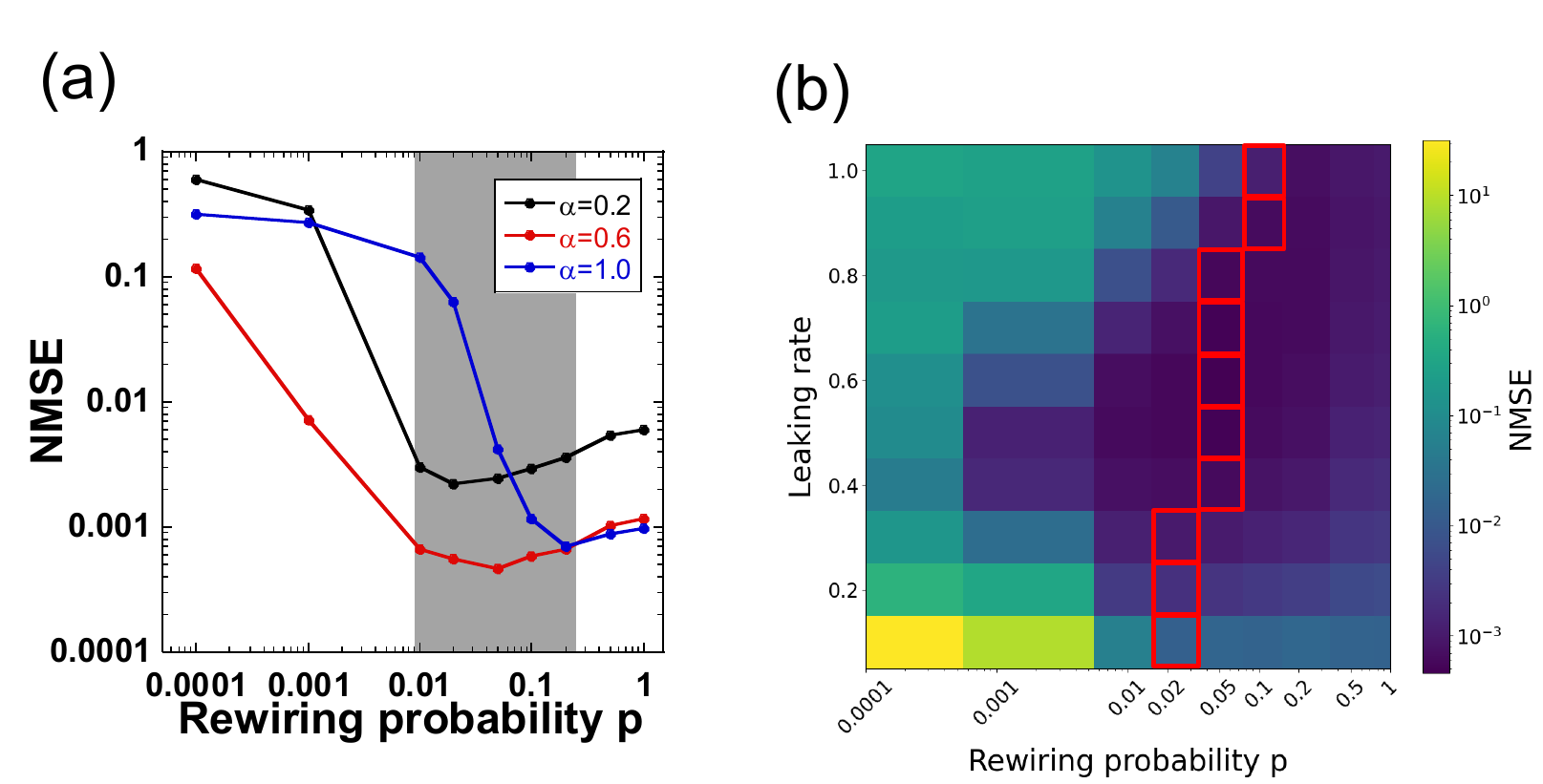}
\caption{NMSE of the time-series prediction task by changing the leak rate $\alpha$ and the rewiring probability $p$. (a) NMSE of the time-series prediction task when varying $p$ at $\alpha = 0.2$ (black line), $\alpha = 0.6$ (red line), and $\alpha = 1.0$ (blue line). (b) Two-dimensional map of the NMSEs as functions of $\alpha$ and $p$. Red squares indicate the optimal conditions for fixed $\alpha$ by changing $p$.}
\label{fig:fig8}
\end{figure}

\subsection{Parameter dependence for time-series prediction}
Next, we evaluate the parameter dependence of $\alpha$ and $p$ on the prediction performance (NMSE) using the one-step prediction task of the Mackey-Glass chaotic time series. Figure~\ref{fig:fig8}(a) shows the changes in the NMSE for different $\alpha$. We found that the optimal $p$ shifts to the larger side as $\alpha$ increases. This suggests that the introduction of more shortcuts into the network enhances the mixing of node states and information propagation for a large leak rate, which improves the performance of time-series prediction. Figure~\ref{fig:fig8}(b) shows the two-dimensional map of the NMSEs when $\alpha$ and $p$ are changed simultaneously. The prediction performance also strongly depends on the network topology, similar to the MC, and the region where small NMSEs appear is located in the small-world region ($0.01 \le p \le 0.2$). However, the optimal condition (red squares) that minimizes the NMSE does not necessarily match the condition that maximizes the MC. Although a large MC is an advantageous factor for the prediction performance, the performance cannot be determined only by the MC because the prediction task requires a balance among the nonlinear mapping, state mixing, and short- and medium-term memory. From these results, we confirm that the network topology and its parameter values in the reservoir needs to be optimized for time-series prediction.

\section{Photonic human brain network as a reservoir}
\subsection{Introduction of the brain functional network using connectomes}
In this section, we evaluate the performance of time-series prediction when a human brain functional network is used as the connection topology of the reservoir. A brain functional network can be estimated from the simultaneous activity (correlation) between brain regions. The global connection structure of brain activity can be treated as a graph by representing nodes as brain regions and edges as functional connections between regions. Recently, the studies that introduce such brain networks into reservoir computing have been reported to analyze the effect of network topologies derived from real-world data on computing performance \cite{Suarez2024}.

We construct a connection matrix (adjacency matrix) for the reservoir from brain functional connectivity data. We use \texttt{conn2res} as a toolbox to connect the nodes in the reservoir \cite{Suarez2024}. The \texttt{conn2res} is a framework that converts brain functional connectivity matrices (connectomes) into reservoir connection matrices and handles the evaluation of reservoir computing consistently. This enables us to utilize human brain networks as reservoir topologies for the comparison of prediction performance.

The human-brain functional network used in this study consists of 1015 nodes. Among them, 1000 nodes belong to the cerebral cortex (ctx), while the remaining 15 nodes correspond to subcortical regions (subctx). The cortical nodes are further classified into seven functional networks: the Dorsal Attention Network (DA, 58 nodes), Default Mode Network (DMN, 277 nodes), Frontoparietal Control Network (FP, 64 nodes), Limbic Network (LIM, 101 nodes), Somatomotor Network (SM, 219 nodes), Ventral Attention Network (VA, 120 nodes), and Visual Network (VIS, 151 nodes). Together with the subcortical regions (subctx, 15 nodes), these constitute the eight brain-network categories represented by different colors in Figs. 9(a) and 9(b). The edges represent connections defined by the brain connectivity matrix used to construct the reservoir. In the present implementation, this connectivity matrix is converted into a binary adjacency matrix, so that an element of 1 indicates the presence of a connection between two nodes and an element of 0 indicates its absence. Thus, the reservoir uses only the presence or absence of connections and does not incorporate the original connection-strength information into the reservoir coupling weights.

\begin{figure}[htbp]
\centering
\includegraphics[width=\linewidth]{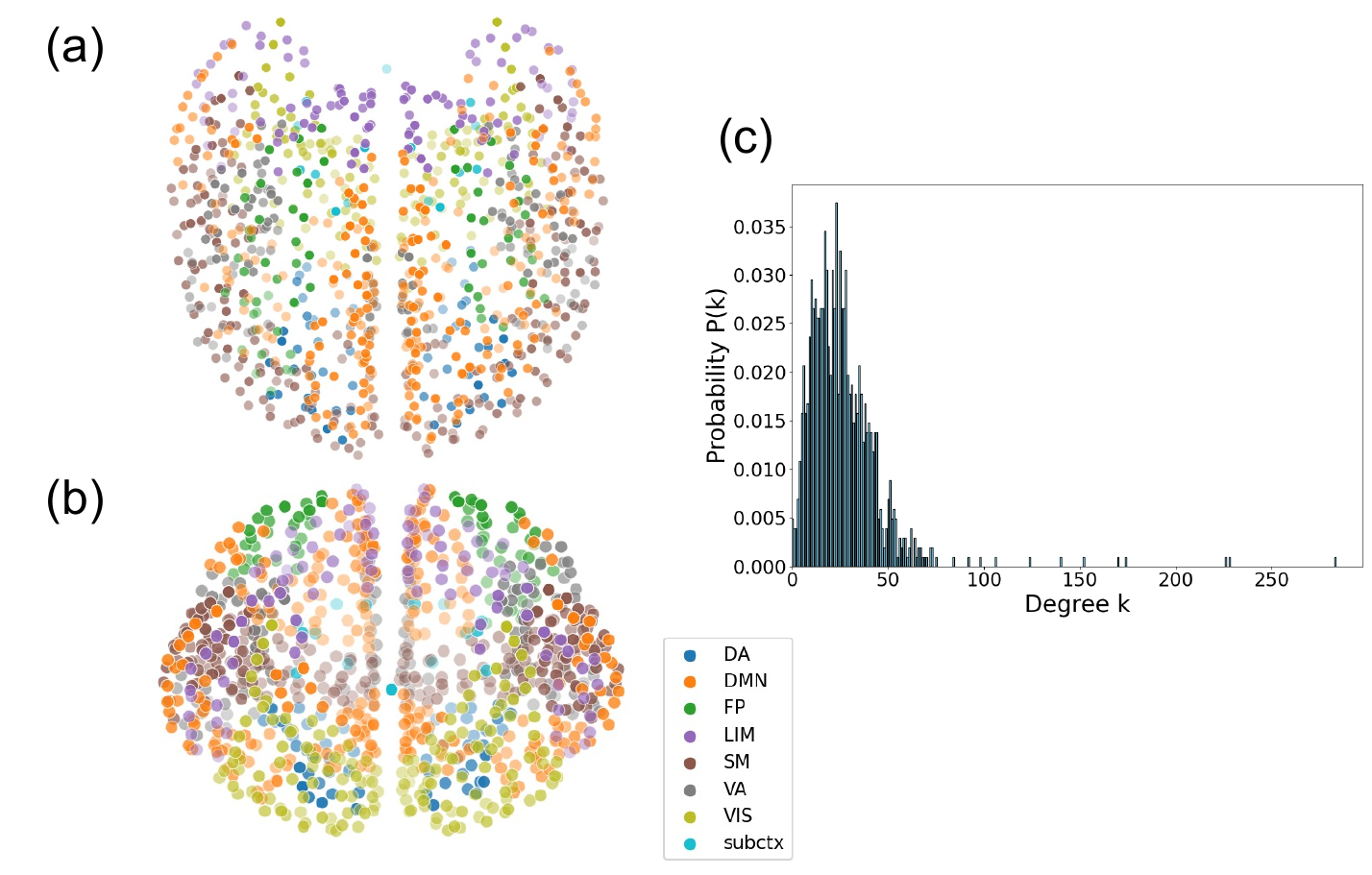}
\caption{Visualization of the human brain functional network from (a) the top and (b) the front. The 1015 nodes are classified into eight brain-network categories, represented by different colors: Dorsal Attention Network (DA, 58 nodes), Default Mode Network (DMN, 277 nodes), Frontoparietal Control Network (FP, 64 nodes), Limbic Network (LIM, 101 nodes), Somatomotor Network (SM, 219 nodes), Ventral Attention Network (VA, 120 nodes), Visual Network (VIS, 151 nodes), and subcortical regions (subctx, 15 nodes). The first seven categories comprise 1000 cortical nodes, while subctx contains 15 subcortical nodes. (c) Degree distribution calculated over all 1015 nodes of the brain network~\cite{Suarez2024}.}
\label{fig:fig9}
\end{figure}

\begin{table}[htbp]
\centering
\setlength{\tabcolsep}{3pt}
\caption{Comparison of network features between the human brain functional network and the WS model.}
\label{tab:ev_brain}
\begin{tabular}{lcccc} 
\toprule
Topology & Average degree & Clustering coefficient & Average path length & Diameter \\
\midrule
Human brain & 26.1 & 0.41 & 2.84 & 6.1 \\
WS model (optimized)    & 8.00 & 0.47 & 5.00 & 9.0 \\
\bottomrule
\end{tabular}
\end{table}

The degree distribution in Fig. 9(c) is calculated over all 1015 nodes of the brain network. The average degree of the entire network is 26.1, indicating that each node is connected to approximately 26 other nodes on average. The degree is not uniform across the different brain-network categories; for example, the average degree calculated within each category varies among the functional networks. Therefore, Fig. 9(c) represents the global degree distribution of the entire brain network rather than separate degree distributions for individual brain regions. The degree distribution appears to have the characteristics of both the regular degree distribution of a small-world network and a scale-free network. It has been reported that brain functional networks possess small-world properties~\cite{Stam2004} (i.e., achieving both a high clustering coefficient and a short average path length). 

Table ~\ref{tab:ev_brain} shows the evaluation results of the brain functional network used in this study. The human brain network exhibits a large clustering coefficient and a short average path length, which are the characteristic of a small-world network.

In the evaluation using the brain functional network, we use the one-step prediction task of the Mackey-Glass chaotic time series, and we use the NMSE as the performance metric. To examine the applicability of the present photonic reservoir framework to a real-world network topology, we evaluate the human-brain functional network and compare its prediction performance with that of an artificial small-world network generated by the WS model. This comparison is intended to provide a reference within the present photonic reservoir framework. The human-brain network contains 1015 nodes, which is close in size to the 1024-node WS reservoir used for comparison. The training data length and regularization coefficient are kept the same for the performance comparison.

\begin{figure}[htbp]
\centering
\includegraphics[width=\linewidth]{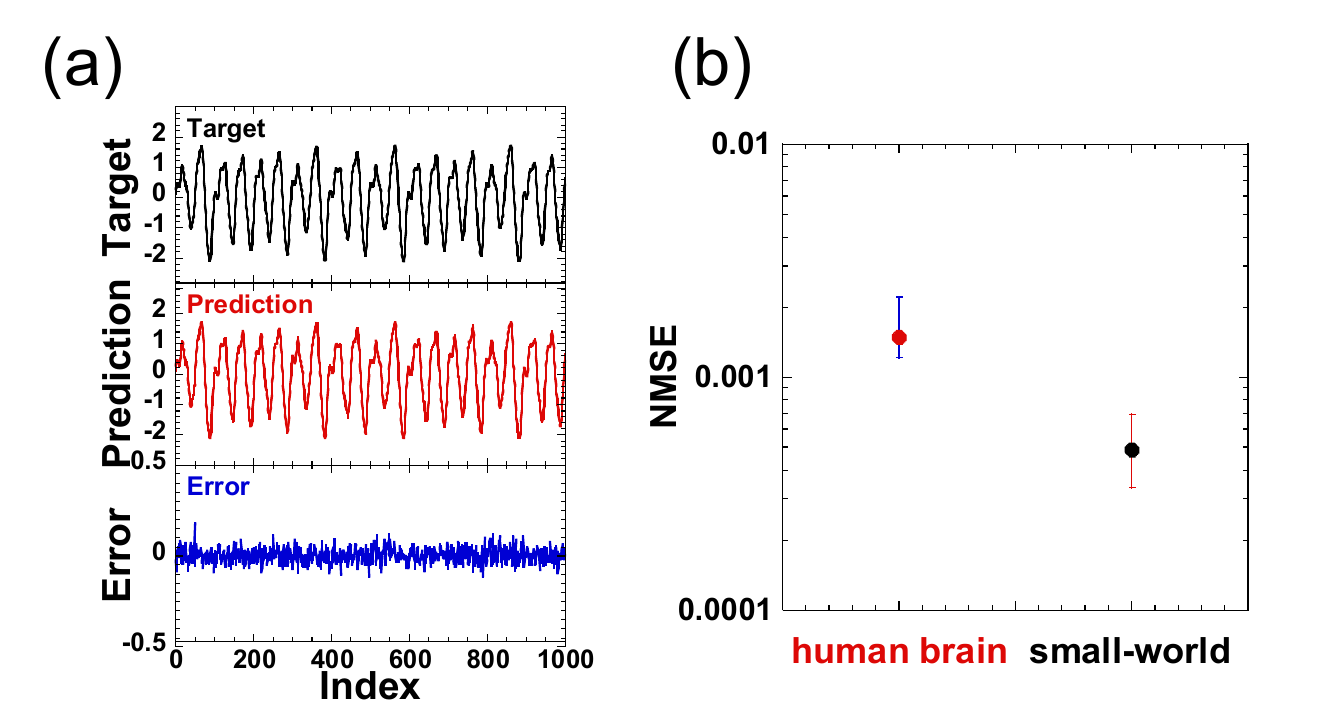}
\caption{(a) Temporal waveforms of the target signal, the prediction signal, and the error between them when the human brain functional network is used. (b) Comparison of the NMSE between the brain functional network and the small-world network. The error bars indicate the results of ten trials.}
\label{fig:fig10}
\end{figure}

\subsection{Results of the photonic human brain network}
Figure~\ref{fig:fig10}(a) shows the temporal waveforms of the target signal, the prediction signal, and the prediction error for the one-step-ahead prediction using the human brain functional network. We confirm that the prediction signal follows the target signal, and a certain degree of accuracy is obtained as a short-term prediction. This result indicates that the topology derived from real human-brain data can be effective as a reservoir.

Figure~\ref{fig:fig10}(b) shows the comparison of the NMSE between the human brain functional network and the WS small-world network adjusted for the task. The optimized WS small-world network achieves a lower NMSE, although the brain functional network exhibits a certain level of prediction performance. Therefore, possessing small-world properties can contribute to the prediction performance. It is important that the network structure is adapted to the task and the dynamical conditions.

Notably, while the human brain functional network exhibits an even shorter average path length ($2.84$) than the optimized WS model ($5.00$), this high propagation efficiency alone does not guarantee superior prediction performance.
In other words, the prediction performance is not determined solely by the "type" of topology (whether it is small-world or not), however, is optimized by the combination of the dynamical parameters and task characteristics. We consider that the adjustable topology, such as the WS model, has the advantage for optimization, whereas the brain functional network has a smaller adjustability of the network topology because it is derived from real-world data. We consider that this difference may appear as the difference in prediction performance.

We note that the human-brain functional network and the WS network differ in basic graph properties such as connection density and degree distribution. In the present study, we did not perform calculations using randomized surrogate networks matched to these properties or to the spectral characteristics of the connectome. Such comparisons will be important in future work to clarify which structural properties of real-world networks contribute to reservoir performance.

\section{Discussions}
We experimentally and numerically investigated the influence of complex-network topology on a large-scale SLM-based photonic reservoir. The present results should be distinguished from the general observation that reservoir topology can affect computational performance, which has already been demonstrated in previous numerical studies \cite{Kawai2019,Haluszczynski2019,Dale2021,Kitayama2022}. Earlier studies have shown that small-world, scale-free, random, and other structured reservoirs can exhibit different computational properties and that the relative advantage of a particular topology can depend on the reservoir configuration, target system, and computational task.

The significance of the present results lies in the systematic implementation and comparison of different complex-network topologies on the experimental physical photonic reservoir platform. In particular, the experimentally observed dependence of the memory capacity and prediction performance on the network topology is qualitatively reproduced by the corresponding numerical simulations. This experiment--simulation correspondence indicates that topology-dependent characteristics investigated in numerical reservoir models can also emerge in a physical large-scale photonic implementation.

The agreement between the experimental and numerical results also clarifies the role of the photonic implementation. Our experiment demonstrates that topology-dependent computational behavior predicted by the numerical model can be physically realized in a large-scale spatially multiplexed photonic reservoir under practical experimental conditions, including optical noise, finite spatial and intensity resolution, alignment errors, and device imperfections.

A key hardware advantage of the present photonic implementation is the spatial parallelism provided by light. The SLM represents 1024 physical reservoir nodes as spatial macropixels, and their nonlinear optical transformation and spatial state generation are performed in parallel. However, the network coupling, feedback calculation, and output training and readout are digitally implemented in the present system. Therefore, the current setup should be regarded as a hybrid optical--digital implementation. Further implementation of the network coupling and readout in the optical domain would allow the spatial parallelism of light to be exploited more extensively and is an important direction for future work.

To further examine whether the performance difference can be attributed to spectral properties of the reservoir coupling matrix, we additionally compared the small-world network with a spectrum-matched control network. The control coupling matrix was constructed to have the same eigenvalue spectrum, and hence the same spectral radius, as the small-world-network coupling matrix (see Appendix for details). Despite the identical eigenvalue spectrum, the small-world network achieved an NMSE of $3.37\times10^{-4}$, whereas the spectrum-matched control network achieved an NMSE of $1.40\times10^{-3}$ for the chaotic time-series prediction task. These results indicate that the performance difference cannot be explained solely by the spectral radius or, more generally, by the eigenvalue spectrum of the coupling matrix. Therefore, although variations in the rewiring probability may affect both network structure and spectral properties, structural properties beyond the eigenvalue spectrum also contribute to the reservoir performance.

We also note that the random-network limit of the WS model at $p=1$ serves as a baseline for evaluating the small-world topology under the controlled binary-weight condition used throughout this study. The WS model continuously changes from a regular network at $p=0$, through the small-world regime at intermediate $p$, to a random network at $p=1$. As shown in Figs. 5 and 6, favorable performance is obtained in the small-world region compared with the random-network regime. Thus, the advantage observed in the small-world region is not only relative to the regular network but also to a random network constructed under the same binary-weight condition.

The topology-dependent computational performance observed in this study can be interpreted in terms of the structural properties of the reservoir networks. In particular, the small-world topology combines relatively high local clustering with short path lengths resulting from a limited number of long-range connections. The high local clustering can maintain recurrent interactions among neighboring reservoir nodes, whereas the short path lengths can facilitate the propagation and mixing of input-dependent states across the network. The coexistence of these two structural properties may therefore provide a balance between local information retention and global information transfer, which offers a possible explanation for the improved memory and prediction performance observed for the small-world networks under the conditions investigated here. However, this interpretation is based on the network properties and computational performance evaluated in the present study. A more direct analysis of information propagation through the reservoir and of the distribution of trained readout weights for different network topologies would provide further insight into the underlying mechanism and represents an important direction for future investigation.

A conventional reservoir is also often constructed using sparse random connections with randomly assigned signed and weighted values. Such a reservoir represents a different baseline from the binary random network considered here. Introducing random signed and weighted connections would change both the network structure and the weight distribution, making it difficult to isolate their respective contributions within the present comparison. Therefore, the present study focuses on controlled comparisons of network topology using binary coupling matrices. A systematic investigation of the combined effects of network topology and connection-weight distributions, including signed and analog-valued weights, remains an important direction for future work.

Our results do not imply that the small-world topology is universally optimal. In our experiments, the rewiring probability maximizing the memory capacity ($p=0.2$) differs from that minimizing the prediction error ($p=0.05$). This task dependence is consistent with previous numerical studies showing that the relative performance of different reservoir topologies can vary with the computational task and system dynamics \cite{Kawai2019,Haluszczynski2019}. Therefore, the network topology should be considered jointly with the reservoir dynamical parameters and the target task rather than optimized independently.

From this viewpoint, the small-world region provides a useful design space for the present SLM-based photonic reservoir rather than a universally optimal network structure. Our numerical results further show that the favorable rewiring probability changes with the leak rate, demonstrating the importance of jointly optimizing structural and dynamical parameters. These results provide experimentally supported design guidelines for topology design in large-scale photonic reservoirs. Thus, network topology can be treated as an additional design parameter, together with conventional reservoir dynamical parameters, when optimizing large-scale photonic reservoirs for a specific computational task.\par

Future works include (i) implementation of the reservoir network connectivity in the optical domain, for example using an additional SLM or a diffractive optical element, to further exploit the spatial parallelism of light while addressing practical issues such as alignment, scalability, stability, and signal-to-noise ratio; (ii) refinement of the design guidelines by considering the effects of optical noise and quantization; and (iii) adaptive, task-dependent optimization of the network topology, including automated optimization or rewiring of the network connections in response to the computational task.

\section{Conclusions}
In this study, we systematically investigated the effect of the internal connection structure (network topology) on the performance of photonic reservoir computing using a SLM in both experiment and numerical simulations. We constructed a large-scale reservoir by regarding macropixels on the SLM as nodes and established a framework to control the connection structure using the adjacency matrices of complex networks. Furthermore, we confirmed that the optical spatiotemporal dynamics exhibit similar bifurcation structures in both the experimental and numerical results, indicating that the experimental system operates as a nonlinear system that can be well reproduced by the numerical model.

In the experiment, we changed the network topology using the WS and BA models and evaluated the performance using the MC and the NMSE of the chaotic time series prediction. We found that both the MC and time-series prediction performance are improved in the small-world region of the WS model under the experimental conditions and tasks considered in this study. We also confirm the validity of the experimental system (model consistency) as the numerical calculations under the same conditions show a similar tendency. On the contrary, the BA model shows lower performance than the WS model, indicating that the scale-free structure is not necessarily advantageous under the conditions of this study.

In the numerical calculations, we evaluated the MC and the NMSE when the leak rate $\alpha$ and the rewiring probability $p$ are changed to investigate the relationship between the topology and the dynamical parameters. For the MC, we found that the maximum value is achieved in the small-world region ($0.01 \le p \le 0.2$) for all $\alpha$, and the MC tends to increase as $\alpha$ increases. Regarding the time-series prediction performance, high performance is obtained in the small-world region. However, the optimal rewiring probability $p$ that minimizes the NMSE depends on $\alpha$. This suggests that the topology needs to be optimized by the parameter change in the reservoir.

Furthermore, we introduced a human brain functional network as the reservoir topology using connectomes and evaluated its time-series prediction performance. We found that while the human brain functional network exhibits metrics of small-worldness, the artificial small-world network (WS model) adjusted for the task can achieve higher performance. These results indicate that while "small-worldness" is an effective guideline, it is not a sufficient condition for achieving the best performance. It is crucial to design a topology that is consistent with the task and dynamical conditions.

These results indicate that small-worldness can provide a useful design guideline for the present photonic reservoir, while the optimal network structure depends on the computational task and reservoir dynamical conditions. These results demonstrate that network topology can be treated as an additional design parameter, together with reservoir dynamical parameters, for improving large-scale photonic reservoir computing.

\section{Appendix: Spectrum-matched control analysis}
To distinguish the contribution of the eigenvalue spectrum from other structural properties of the reservoir coupling matrix, we performed an additional numerical analysis by matching the eigenvalue spectrum between a small-world-network coupling matrix and a coupling matrix derived from a random network.

Let the coupling matrices of the small-world and random networks be denoted by $W_{\mathrm{SW}}$ and $W_{\mathrm{R}}$, respectively. Their eigendecompositions are written as
\[
W_{\mathrm{SW}}=P_{\mathrm{SW}}\Lambda_{\mathrm{SW}}P_{\mathrm{SW}}^{-1},
\]
and
\[
W_{\mathrm{R}}=P_{\mathrm{R}}\Lambda_{\mathrm{R}}P_{\mathrm{R}}^{-1},
\]
where $P_{\mathrm{SW}}$ and $P_{\mathrm{R}}$ contain the corresponding eigenvectors, and $\Lambda_{\mathrm{SW}}$ and $\Lambda_{\mathrm{R}}$ are diagonal matrices containing the corresponding eigenvalues.

We then constructed a spectrum-matched control coupling matrix as
\[
W_{\mathrm{control}}=P_{\mathrm{R}}\Lambda_{\mathrm{SW}}P_{\mathrm{R}}^{-1}.
\]
In this construction, the eigenvector basis $P_{\mathrm{R}}$ is obtained from the random-network coupling matrix, whereas the eigenvalues are replaced by those of the small-world-network coupling matrix. Consequently, $W_{\mathrm{control}}$ and $W_{\mathrm{SW}}$ have the same eigenvalue spectrum and therefore exactly the same spectral radius.

\begin{figure}[htbp]
\centering
\includegraphics[width=0.95\linewidth]{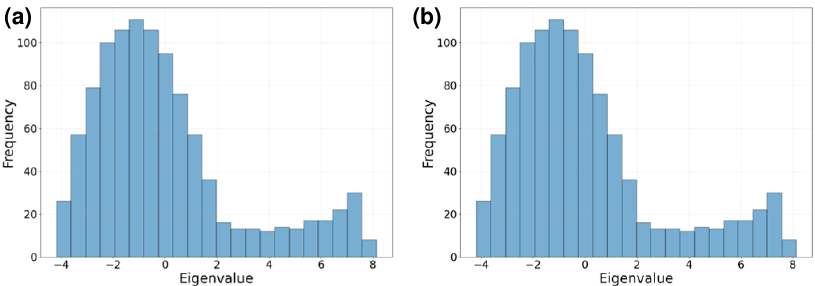}
\caption{Eigenvalue distributions of (a) the small-world network and (b) the spectrum-matched random network.}
\label{fig:appendix_eigdist}
\end{figure}

\begin{figure}[htbp]
\centering
\includegraphics[width=0.95\linewidth]{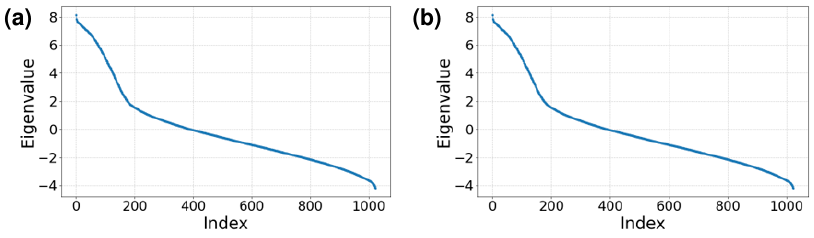}
\caption{Eigenvalues arranged in descending order for (a) the small-world network and (b) the spectrum-matched random network.}
\label{fig:appendix_eigorder}
\end{figure}

The agreement of the eigenvalue spectra is confirmed by comparing the eigenvalue distributions and the eigenvalues arrange in descending order. Figure~\ref{fig:appendix_eigdist} compares the eigenvalue distributions of the spectrum-matched random and small-world networks, while Fig.~\ref{fig:appendix_eigorder} shows the corresponding eigenvalues arranged in descending order. The two coupling matrices exhibit identical eigenvalue spectra, while their eigenvector bases are different.

\begin{figure}[htbp]
\centering
\includegraphics[width=0.55\linewidth]{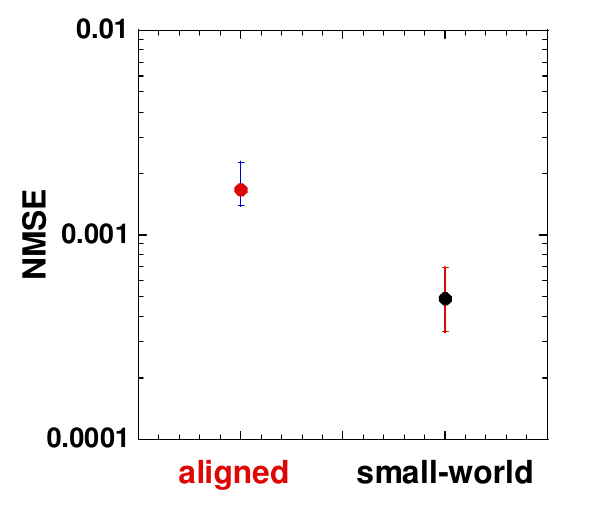}
\caption{Prediction performance (NMSE) of the spectrum-matched random network (aligned) and the small-world network. Error bars indicate the results over 10 trials.}
\label{fig:appendix_nmse}
\end{figure}

We evaluate these two coupling matrices using the same chaotic time-series prediction task, as shown in Fig.~\ref{fig:appendix_nmse}. The small-world network achieves an NMSE of $3.37\times10^{-4}$, whereas the spectrum-matched random network achieves an NMSE of $1.40\times10^{-3}$. The comparison is repeated over 10 trials, and the performance difference is consistently observed.

Their spectral radii are also identical, because the two coupling matrices have the same complete eigenvalue spectrum. Therefore, the observed performance difference cannot be explained by the spectral radius or the eigenvalue spectrum alone. This control analysis suggests that properties beyond the eigenvalue spectrum, including differences associated with the eigenvector basis and network structure, contribute to the reservoir performance.

\section*{Funding} Japan Society for the Promotion of Science (JP22H05195, JP25H01129); CREST Japan Science and Technology Agency (JPMJCR24R2).

\section*{Disclosures} The authors declare no conflicts of interest.

\section*{Data availability} Data underlying the results presented in this paper are not publicly available at this time but may be obtained from the authors upon reasonable request.

\bibliography{export}

\end{document}